\documentclass[12pt]{article}
\usepackage{amsmath}
\usepackage{amsfonts}
\usepackage{amssymb}
\usepackage{graphicx}
\usepackage[left=2cm,right=2cm,top=2cm,bottom=2cm]{geometry}
\usepackage[sort&compress,square,numbers,comma]{natbib}
\usepackage[colorlinks=true,citecolor=blue,filecolor=magenta]{hyperref}

\usepackage{soul}
\usepackage{changes}
\definecolor{brown-web}{rgb}{0.65, 0.16, 0.16}
\setaddedmarkup{\textcolor{red}{\textbf{#1}}}
\setdeletedmarkup{\textcolor{brown-web}{\sout{#1}}}

\title{MECHANICAL ANALOGY FOR THE WAVE OF NUCLEAR BURNING}

\author{V.V. Urbanevich, I.V. Sharph, V.A. Tarasov, V.D. Rusov}

\date{}

\begin{document}

\maketitle

\begin{center}
     \textit{Odessa National Polytechnic University,\\
     \mbox{~~}Shevchenko av. 1, Odessa 65000, Ukraine}


\end{center}


\begin{abstract}
We consider a model of neutron-nuclear wave burning.
The wave of nuclear burning of the medium is initiated by an
external neutron source and is the basis for the new generation reactors -- the
so-called "traveling-wave reactors".


We develop a model of nuclear wave burning, for which it is possible to draw an
analogy with a mechanical dissipative system. Within the framework of the new
model, we show that two burning modes are possible depending on the control
parameters: a traveling autowave and a wave driven by an external neutron
source. We find the autowave to be possible for certain neutron energies only,
and the wave velocity has a continuous spectrum bounded below.
\end{abstract}


\section{Introduction}
A problem of providing the humankind with energy has been around for a while.
Today the experts tend to associate its solution with two major directions:
the fusion reactors and the $5^{th}$ generation nuclear reactors (as well as 
their hybrid versions -- fusion-fission reactors)~\citep{1,2,3,4,5,6,7,8}. And
today, while these projects are not yet implemented in practice, their
importance largely depends on which one will be created first.

The present paper is devoted to the theoretical study of the wave 
neutron-nuclear burning modes, which are the basis for some nuclear reactors of
the $5^{th}$ generation (Gen-V), e.g.~\citep{1,2,3,4,9,10,11,12,13,14,15,16,17,
18,19,20,21,22,23,24,25,26,27,28,29,30,31,32,33,34,35,36,37,38,39,40,41}. 
These reactors are often referred to as the Feoktistov reactors (in USA more 
common "Traveling-wave reactors", and in Japan -- "CANDLE"-reactors). In our 
opinion, these reactors are the most promising among the Gen-V reactors. In 
contrast to the previous generation reactors, they do not require the 
super-critical fuel load. Therefore they basically cannot explode as a result
of uncontrolled fission chain reaction, which classifies them as safe
reactors~\citep{1}. At the same time they involve a non-linear self-regulating
neutron-fission wave of slow nuclear burning that does not require a human
intervention to regulate reactivity. This puts them into a class of even more 
safe reactors -- the reactors with inherent safety.

Such reactors may have a variety of technical design features depending on
their purpose, but the primary criterion to classify them as reactors with
inherent safety is the the implementation of the wave mode of nuclear burning.

One can also imagine these reactors to be implemented as the transmutation 
reactors (biocompatibility)~\citep{3,32,41,42}. For example, the 
uranium-plutonium reactor that operates on the wave nuclear burning with 
intermediate neutrons.

In the present paper we study a number of features of the Feoktistov reactor
concept. The major one is that the composition and structure of the reactor 
core, as well as the external parameters, must be carefully picked in such a
way that they satisfy two conditions. First, its characteristic time must be
much more than a typical time of chain neutron reaction, which is the average
lifetime of a neutron generation, and which is mainly determined by the average
time of delayed neutrons emission. Second, some elements of self-regulation 
must appear in this mode~\citep{1,9}.

This can be achieved if the following chain of transformations is dominant 
among the nuclear reactions in the core:
	
\begin{equation}
{}_{92}^{238}U + n \to {}_{92}^{239}U \to {}_{93}^{239}Np \to {}_{94}^{239}Pu.
\label{1}
\end{equation}

In this case the plutonium, produced by this chain of transformations is the
main fuel. The characteristic time of this reaction is roughly the time of two
beta decays, which is about $\tau_{\beta} = \frac{2.3}{\ln (2)} = 3.3$ days. It
is almost four orders of magnitude larger than the typical time for delayed
neutrons. There is also a similar thorium-uranium chain of transformations,
e.g.~\citep{10}.

Although the basic kinetic model of the neutron-nuclear burning wave is 
extensively studied, there is still very little information about some of its
features and the possibilities of its implementation, e.g.~\citep{29}.

Almost nothing is known about how the burning modes depend on the
characteristics of an external source of neutrons. The requirements for the 
neutron source are therefore not defined as well. Very little is known about
the kinetics of the steady burning wave formation and about the wave velocity,
which determines the reactor's heat power. The impact of heat transfer and
neutron spectrum, the possible composition of the fuel, its structure and phase
state etc. are all undefined.

\section{System of kinetic equations}
\label{section1}

Let us start with the system of kinetic (balance) equations for neutrons and
nuclides, which describe the process of wave neutron-nuclear burning of 
uranium-plutonium fissile medium. We use the work by L.P.~Feoktistov~\citep{9} 
as a basis. Note that the system considered further on, has a simplified form.
As in~\citep{9}, the equations describing the fission fragments are omitted,
and the fuel is considered initially non-enriched, and consisting of $^{238}U$
only (i.e. there are no fissile nuclides like $^{235}U$, as in~\citep{32}).

Let us first consider the equation for neutrons. With the mentioned
simplifications in the diffusion approximation it looks as follows:

\begin{eqnarray}
   \frac{\partial{ \tilde{n}\left( t,x \right)}}{\partial t}=D\frac{{{\partial }^{2}}{\tilde{n}\left( t,x \right)}}{\partial {{x}^{2}}}+\left( \nu -1 \right)\tilde{n}\left( t,x \right)\left| { \vec {\mathbf{v}} } \right|{{N}_{Pu}}\left( t,x \right){{\sigma }_{f,Pu}}- \nonumber \\
   -\tilde{n}\left( t,x \right)\left| \vec {\mathbf{v}} \right|\left( {{N}_{8}}\left( t,x \right){{\sigma }_{a,8}}+{{N}_{9}}\left( t,x \right){{\sigma }_{a,9}}+{{N}_{Pu}}\left( t,x \right){{\sigma }_{c,Pu}} \right), 
\label{2}
\end{eqnarray}

\noindent
where $\tilde{n}$, ${N_{8}}$, ${N_{9}}$ and ${N_{Pu}}$ are the concentrations
of neutrons, $^{238}U$, $^{239}U$ $^{239}Pu$ respectively; ${{\sigma }_{c,i}}$
is the neutron capture cross-section for the $i^{th}$ nuclide, 
${{\sigma }_{f,i}}$ is the fission cross-section, ${{\sigma }_{a,i}}$ is the
neutron absorption cross-section 
(${{\sigma }_{a,i}}={{\sigma }_{f,i}}+{{\sigma }_{c,i}}$); ${D}$ is the neutron
diffusion coefficient; $\nu $ is the average number of neutrons produced per
$^{239}Pu$ fission.

For further convenience we introduce the dimensionless concentrations:

\begin{equation}
n\left( {t,x} \right) = \frac{{\tilde{n}\left( {t,x} \right)}}{{N_8^0}},{n_{Pu}}\left( {t,x} \right) = \frac{{{N_{Pu}}\left( {t,x} \right)}}{{N_8^0}},{n_8}\left( {t,x} \right) = \frac{{{N_8}\left( {t,x} \right)}}{{N_8^0}},{n_9}\left( {t,x} \right) = \frac{{{N_9}\left( {t,x} \right)}}{{N_8^0}},
\label{3}
\end{equation}

\noindent
where $N^{8}_{0}$ is the initial concentration of $^{238}U$.

With these dimensionless quantities, the equation (\ref{2}) will take the form:

\begin{eqnarray}
\frac{{\partial n\left( {t,x} \right)}}{{\partial t}} = D\frac{{{\partial ^2}n\left( {t,x} \right)}}{{\partial {x^2}}} + \left( {\left( {N_8^0\left| {\vec {\mathbf{v}}} \right|{\sigma _{f,Pu}}} \right)\left( {\nu  - 1} \right) - (N_8^0\left| {\vec {\mathbf{v}}} \right|{\sigma _{c,Pu}}}) \right)n\left( {t,x} \right){n_{Pu}}\left( {t,x} \right)-\nonumber \\
 - n\left( {t,x} \right)\left( {{n_8}\left( {t,x} \right)\left( {N_8^0\left| {\vec {\mathbf{v}}} \right|{\sigma _{a,8}}} \right) + {n_9}\left( {t,x} \right)\left( {N_8^0\left| {\vec {\mathbf{v}}} \right|{\sigma _{a,9}}} \right)} \right).
\label{2.1}
\end{eqnarray}

$N_{8}^{0}\left| {\vec {\mathbf{v}}} \right|{{\sigma }_{f,Pu}}$,
$N_{8}^{0}\left| {\vec {\mathbf{v}}} \right|{{\sigma }_{c,Pu}}$,
$N_{8}^{0}\left| {\vec {\mathbf{v}}} \right|{{\sigma }_{a,8}}$, and 
$N_{8}^{0}\left| {\vec {\mathbf{v}}} \right|{{\sigma }_{a,9}}$ have the
dimension of inverse time. I.e. these quantities represent the inverse mean
free times for the neutrons with respect to the corresponding nuclear reaction
with relevant nucleus, given the nuclei concentrations $N^{8}_{0}$.
Therefore, they may be denoted as follows:

\begin{equation}
N_{8}^{0}\left| {\vec {\mathbf{v}}} \right|{{\sigma }_{f,Pu}}=\frac{1}{{{\tau }_{f,Pu}}},N_{8}^{0}\left| {\vec {\mathbf{v}}} \right|{{\sigma }_{c,Pu}}=\frac{1}{{{\tau }_{c,Pu}}},N_{8}^{0}\left| {\vec {\mathbf{v}}} \right|{{\sigma }_{a,8}}=\frac{1}{{{\tau }_{a,8}}},N_{8}^{0}\left| {\vec {\mathbf{v}}} \right|{{\sigma }_{a,9}}=\frac{1}{{{\tau }_{a,9}}}.
\label{4}
\end{equation}

Than equation (\ref{2.1}) then becomes

\begin{eqnarray}
	\frac{{\partial n \left( {t,x} \right)}}{{\partial t}} = D\frac{{{\partial ^2}n \left( {t,x} \right)}}{{\partial {x^2}}} + \left( {\frac{1}{{{\tau _{f,Pu}}}}\left( {\nu  - 1} \right) - \frac{1}{{{\tau _{c,Pu}}}}} \right)n\left( {t,x} \right){n_{Pu}}\left( {t,x} \right) -\nonumber \\
	- n\left( {t,x} \right)\left( {{n_8}\left( {t,x} \right)\frac{1}{{{\tau _{a,8}}}} + {n_9}\left( {t,x} \right)\frac{1}{{{\tau _{a,9}}}}} \right).
\label{5}
\end{eqnarray}

However, the introduced mean free times are approximately equal. At least, the
difference between them is much less than their difference from the overall time
scale of the problem -- the characteristic time of $\beta$-decay
${{\tau }_{\beta }}$. Taking an approximation
 
\begin{equation}
	{{\tau }_{f,Pu}}\approx{{\tau }_{c,Pu}}\approx {{\tau }_{a,8}}\approx {{\tau }_{a,9}}=\tau, 
\label{6}
\end{equation}

\noindent	
and taking it into account in the equation (\ref{5}), we obtain the following
expression:

\begin{equation}
	\frac{\partial n \left( {t,x} \right)}{\partial t}=D\frac{{{\partial }^{2}}n \left( {t,x} \right)}{\partial {{x}^{2}}}+\frac{1}{\tau }\left( \left( \nu -2 \right){{n}_{Pu}}\left( t,x \right)-{{n}_{8}}\left( t,x \right)-{{n}_{9}}\left( t,x \right) \right)n\left( t,x \right).
\label{7}
\end{equation}

Let us now switch from the time $t$ to a new dimensionless time, which we also
denote $t$ and which is equal to the old $t$ divided by $\tau$. We also
introduce a dimensionless coordinate:

\begin{equation}
x=\sqrt{D\tau }y.
\label{8}
\end{equation}

Then the neutron equation (\ref{7}) will look like:

\begin{equation}
	\frac{\partial n \left( t,y \right)}{\partial t}=\frac{{{\partial }^{2}}n \left( t,y \right)}{\partial {{y}^{2}}}+\left( \left( \nu -2 \right){{n}_{Pu}}\left( t,y \right)-{{n}_{8}}\left( t,y \right)-{{n}_{9}}\left( t,y \right) \right)n\left( t,y \right)\\.
\label{9.1}
\end{equation}

We suppose that $^{238}U$ can only burn out and cannot accumulate in any way.
Then its kinetic equation is

\begin{equation}
 \frac{\partial {{N}_{8}}\left( t,x \right)}{\partial t}=- \sigma_{a,8} \left| {\vec {\mathbf{v}}} \right| {{N}_{8}}\left( t,x \right)\tilde{n}\left( t,x \right).
\label{9.1.1}
\end{equation}

If we simplify and use the dimensionless values as in (\ref{3}), (\ref{4}),
(\ref{6}) and (\ref{8}), it becomes

\begin{equation}
\frac{\partial {{n}_{8}}\left( t,y \right)}{\partial t}=- {{n}_{8}}\left( t,y \right)n\left( t,y\right).
\label{9.1.2}
\end{equation}

$^{239}U$ is produced during the neutron capture by $^{238}U$. Its amount
decreases due to absorption of neutrons (neutron capture or nuclear fission)
and $\beta$-decay with characteristic time $\tau_{\beta}$. So the equation for
$^{239}U$ is

\begin{equation}
\frac{\partial {{N}_{9}}\left( t,x \right)}{\partial t}= \left( \sigma_{a,8} {{N}_{8}}\left( t,x \right)-\sigma_{a,9}{{N}_{9}}\left( t,x  \right) \right)\left| {\vec {\mathbf{v}}} \right| \tilde{n} \left( t,x \right)-\frac{1}{{{\tau }_{\beta }}}{{N}_{9}}\left( t,x \right).
\label{9.2.1}
\end{equation}

After scaling the equation (\ref{9.2.1}) takes the form:

\begin{equation}
\frac{\partial {{n}_{9}}\left( t,y \right)}{\partial t}=\left( {{n}_{8}}\left( t,y \right)-{{n}_{9}}\left( t,y \right) \right)n\left( t,y \right)-\frac{\tau }{{{\tau }_{\beta }}}{{n}_{9}}\left( t,y \right).
\label{9.2.2}
\end{equation}

Since $^{239}Pu$ is produced by the $\beta$-decay of $^{239}U$, and burns out
absorbing neutrons, the kinetic equation for $^{239}Pu$ is

\begin{equation}
\frac{\partial {{N}_{Pu}}\left( t,x \right)}{\partial t}=\frac{1}{{{\tau }_{\beta }}}{{N}_{9}}\left( t,x \right)-\left(\sigma_{f,Pu}+\sigma_{c,Pu} \right){N}_{Pu}\left( t,x \right) \left| {\vec {\mathbf{v}}} \right| \tilde{n} \left( t,x \right)
\label{9.3.1}
\end{equation}

After simplification and scaling as in (\ref{3}), (\ref{4}), (\ref{6}) and
(\ref{8}), it will take the form:

\begin{equation}
 \frac{\partial {{n}_{Pu}}\left( t,y \right)}{\partial t}=\frac{\tau }{{{\tau }_{\beta }}}{{n}_{9}}\left( t,y \right)-\frac{1}{{{{\tilde{n}}}_{Pu}}}{{n}_{Pu}}\left( t,y \right)n\left( t,y \right),
\label{9.3.2}
\end{equation}

\noindent where

\begin{equation}
{\tilde{n}}_{Pu}=\frac{\sigma_{c,8}}{ \sigma_{f,Pu}+\sigma_{c,Pu}}.
\label{9.3.3}
\end{equation}

If we combine the equations (\ref{9.1}), (\ref{9.1.2}), (\ref{9.2.2}) and
(\ref {9.3.2}), we obtain the following system of kinetic equations:

\begin{equation}
	\begin{cases}
   \frac{\partial n}{\partial t}\left( t,y \right)=\frac{{{\partial }^{2}}n}{\partial {{y}^{2}}}\left( t,y \right)+\\
   +\left( \left( \nu -2 \right){{n}_{Pu}}\left( t,y \right)-{{n}_{8}}\left( t,y \right)-{{n}_{9}}\left( t,y \right) \right)n\left( t,y \right), \\ 
  \frac{\partial {{n}_{8}}\left( t,y \right)}{\partial t}=- {{n}_{8}}\left( t,y \right)n\left( t,y\right), \\ 
  \frac{\partial {{n}_{9}}\left( t,y \right)}{\partial t}=\left( {{n}_{8}}\left( t,y \right)-{{n}_{9}}\left( t,y \right) \right)n\left( t,y \right)-\frac{\tau }{{{\tau }_{\beta }}}{{n}_{9}}\left( t,y \right), \\ 
  \frac{\partial {{n}_{Pu}}\left( t,y \right)}{\partial t}=\frac{\tau }{{{\tau }_{\beta }}}{{n}_{9}}\left( t,y \right)-\frac{1}{{{{\tilde{n}}}_{Pu}}}{{n}_{Pu}}\left( t,y \right)n\left( t,y \right). \\ 
	\end{cases}
\label{9}
\end{equation}

As one can see, a very small parameter $\frac {\tau} {{\tau} _{\beta}}$ emerged
in the problem -- the ratio of the neutron mean free time $\tau$ (about
$10^{-7}$~s) to the characteristic time of $\beta$-decay 
${{\tau} _{\beta}} \approx 3~days$. We shall denote this parameter by 
$\varepsilon$, and it is approximately equal to $10^{-14}$~s.

Let us convert the resulting system of kinetic equations into the autowave
form. For this purpose we make the variable substitution according to the
relation $z=y+vt$. With such substitution the burning wave goes in the 
negative direction of the coordinate axis of the autowave variable $z$. The 
derivatives change as follows:

\begin{eqnarray}
\frac{{\partial n}}{{dt}} = v\frac{{\partial n}}{{dz}},\label{10}\\
\frac{{{\partial ^2}n}}{{d{y^2}}} = \frac{{{\partial ^2}n}}{{d{z^2}}}.\label{11}
\end{eqnarray} 

System (\ref{9}) in autowave form reads

\begin{eqnarray}
\begin{cases}
v\frac{{dn}}{{dz}}\left( z \right) = \frac{{{d^2}n}}{{d{z^2}}}\left( z \right) + \left( {\left( {\nu  - 2} \right){n_{Pu}}\left( z \right) - {n_8}\left( z \right) - {n_9}\left( z \right)} \right)n\left( z \right),\\
v\frac{{d{n_8}\left( z \right)}}{{dz}} =  - {n_8}\left( z \right)n\left( z \right),\\
v\frac{{d{n_9}\left( z \right)}}{{dz}} = \left( {{n_8}\left( z \right) - {n_9}\left( z \right)} \right)n\left( z \right) - \varepsilon {n_9}\left( z \right),\\
v\frac{{d{n_{Pu}}\left( z \right)}}{{dz}} = \varepsilon {n_9}\left( z \right) - \frac{1}{{{{\tilde n}_{Pu}}}}{n_{Pu}}\left( z \right)n\left( z \right).
\end{cases}
\label{12}
\end{eqnarray}

The second equation in system (\ref{12}) may be integrated. Than we obtain the
following expression for the $^{238}U$ concentration:

\begin{equation}
{n_8}\left( z \right) = {n_{8, - \infty }}\exp \left( { - \frac{1}{v}\int\limits_{ - \infty }^z {n\left( {{z_1}} \right)d{z_1}} } \right),
\label{14}
\end{equation}

\noindent
where ${n_{8, - \infty }}$ is the $^{238}U$ concentration when 
$\left( z \rightarrow - \infty \right)$. Since we use the values scaled to the
initial $^{238}U$ concentration, ${n_{8, - \infty }}=1$.

Applying the variable substitution and some basic mathematical transformations,
one can integrate the rest of the equations from~(\ref{12}). The system of
equations then takes the form:

\begin{eqnarray}
\begin{cases}
\frac{1}{v}\frac{{dn}}{{dz}}\left( z \right) = n\left( z \right)- {n_9}\left( z \right) - \frac{\varepsilon }{v}\left( {1 + \left( {\nu  - 2} \right){{\tilde n}_{Pu}}} \right)\int\limits_{ - \infty }^z {{n_9}\left( {{z_1}} \right)d{z_1}}+ \\
+ \left( {\nu  - 2} \right){{\tilde n}_{Pu}}{n_{Pu}}\left( z \right) + 2\left( {1 - \exp \left( { - \frac{1}{v}\int\limits_{ - \infty }^z {n\left( {{z_1}} \right)d{z_1}} } \right)} \right),\\
{n_8}\left( z \right) = \exp \left( { - \frac{1}{v}\int\limits_{ - \infty }^z {n\left( {{z_1}} \right)d{z_1}} } \right),\\
{n_9}\left( z \right) = \frac{{1}}{v}\exp \left( { - \frac{1}{v}\int\limits_{ - \infty }^z {n\left( {{z_1}} \right)} d{z_1}} \right)\left( {\exp \left( { - \frac{\varepsilon }{v}z} \right)\int\limits_{ - \infty }^z { n\left( {{z_2}} \right)\exp \left( {\frac{\varepsilon }{v}{z_2}} \right)d{z_2}}} \right),\\
{n_{Pu}}\left( z \right) = \frac{\varepsilon }{v}\exp \left( { - \frac{1}{{v{{\tilde n}_{Pu}}}}\int\limits_{ - \infty }^z {n\left( {{z_1}} \right)d{z_1}} } \right)\int\limits_{ - \infty }^z {{n_9}\left( {{z_2}} \right)\exp \left( {\frac{1}{{v{{\tilde n}_{Pu}}}}\int\limits_{ - \infty }^{{z_2}} {n\left( {{z_1}} \right)d{z_1}}} \right)d{z_2}}.
\end{cases}
\label{15}
\end{eqnarray}

Let us consider the last equation of the system (\ref{15}) in more detail. We 
the direct argument to plus infinity in this expression, i.e. 
$\left( z \rightarrow + \infty \right)$. Then, we have the following:

\begin{multline}
{n_{Pu}}\left( {z \to  + \infty } \right) = \frac{\varepsilon }{v}\exp \left( { - \frac{1}{{v{{\tilde n}_{Pu}}}}\int\limits_{ - \infty }^{ + \infty } {n\left( {{z_1}} \right)d{z_1}} } \right)\times\\
\times\int\limits_{ - \infty }^{ + \infty } {{n_9}\left( {{z_2}} \right)\exp \left( {\frac{1}{{v{{\tilde n}_{Pu}}}}\int\limits_{ - \infty }^{{z_2}} {n\left( {{z_1}} \right)d{z_1}} } \right)d{z_2}}.
\label{16}
\end{multline}

This expression may be conventionally split into three multipliers: the
constant $\frac{\varepsilon}{v}$, the exponent, and the integral. The 
integrand is the exponent multiplied by $^{239}U$ concentration, and the
interval of integration is from $\left(-\infty \right)$ to 
$\left(+ \infty \right)$. The exponent is strictly more than zero throughout
the entire integration interval (the properties of this function). The
$^{239}U$ concentration is a non-negative value, and it has to be greater than
zero at some points, because otherwise there would be no reaction and no wave
of nuclear burning at all. So the integral is of the product of two positive
functions, and consequently, is also a positive quantity which is certainly not
equal to zero. Since the constant (the quotient of two positive values) and the
exponent (property of the function) are also positive values, then the entire
product is certainly more than zero. Thus it may be concluded that the
plutonium cannot burn out completely. Instead, it tends to some constant level:
$n_{Pu}\left (z \to  + \infty  \right) \equiv n_{Pu, + \infty } \ne 0$.
This is also confirmed by the results of numerical simulation of the burning 
kinetics, e.g.~\citep{3,16,29,32,33,34,41}.

Let us assume the following:

\begin{equation}
\int\limits_{ - \infty }^z {d{z_2}} n\left( {{z_2}} \right)\exp \left( {\frac{\varepsilon }{v}{z_2}} \right) \approx n\left( z \right)\int\limits_{ - \infty }^z {d{z_2}} \exp \left( {\frac{\varepsilon }{v}{z_2}} \right).
\label{18}
\end{equation}

This approximation is justified by the fact that the function 
$n \left(z_2 \right)$ in this integral makes the largest contribution near the
upper integration limit. So we take it outside the integral, substituting its
argument by this upper limit.

With this approximation, the system of equations (\ref{15}) will take the form:

\begin{eqnarray}
\begin{cases}
\frac{1}{v}\frac{{dn}}{{dz}}\left( z \right) = n\left( z \right)  - {n_9}\left( z \right) - \frac{\varepsilon }{v}\left( {1 + \left( {\nu  - 2} \right){{\tilde n}_{Pu}}} \right)\int\limits_{ - \infty }^z {{n_9}\left( {{z_1}} \right)d{z_1}}-\\
+ \left( {\nu  - 2} \right){{\tilde n}_{Pu}}{n_{Pu}}\left( z \right) + 2\left( {1 - \exp \left( { - \frac{1}{v}\int\limits_{ - \infty }^z {n\left( {{z_1}} \right)d{z_1}} } \right)} \right),\\
{n_8}\left( z \right) = \exp \left( { - \frac{1}{v}\int\limits_{ - \infty }^z {n\left( {{z_1}} \right)d{z_1}} } \right),\\
{n_9}\left( z \right) = \frac{1}{\varepsilon }\exp \left( { - \frac{1}{v}\int\limits_{ - \infty }^z {n\left( {{z_1}} \right)} d{z_1}} \right)n\left( z \right),\\
{n_{Pu}}\left( z \right) = \frac{{{\tilde n}_{Pu}}}{{1 - {{\tilde n}_{Pu}}}}\left[ {\exp \left( { - \frac{1}{v}\int\limits_{ - \infty }^z {n\left( {{z_1}} \right)} d{z_1}} \right) - \exp \left( { - \frac{1}{{v{{\tilde n}_{Pu}}}}\int\limits_{ - \infty }^z {n\left( {{z_1}} \right)d{z_1}} } \right)} \right].
\end{cases}
\label{17}
\end{eqnarray}

Since the burning wave is propagating in a negative direction, the boundary
conditions are set at minus infinity and have the following form:

\begin{eqnarray}
\begin{split}
n\left( {z \to  - \infty } \right) = 0,{n_8}\left( {z \to  - \infty } \right) = {n_{8, - \infty }=1},\\
{n_9}\left( {z \to  - \infty } \right) = 0,{n_{Pu}}\left( {z \to  - \infty } \right) = 0,
\end{split}
\label{13}
\end{eqnarray}

\noindent
where ${n_{8, - \infty }}$ is the initial $^{238}U$ concentration.

\section{Analogy to Newton's second law}
\label{section2}

Let us consider the first equation of the system (\ref{17}). As we have already
derived the expressions for $^{239}U$ and $^{239}Pu$ from other equations, we
can substitute them here. After substitution we can collect terms and integrate
some parts of this equation. Finally we obtain the following expression:

\begin{multline}
\frac{1}{v}\frac{{dn}}{{dz}}\left( z \right) = \left[ {1 - \frac{1}{\varepsilon }\exp \left( { - \frac{1}{v}\int\limits_{ - \infty }^z {n\left( {{z_1}} \right)} d{z_1}} \right)} \right]n\left( z \right) + \left( {1 - \left( {\nu  - 2} \right){{\tilde n}_{Pu}}} \right)+\\
 + \frac{{(\nu -1) {{\tilde n}_{Pu}} - 1}}{{1 - {{\tilde n}_{Pu}}}}\exp \left( { - \frac{1}{v}\int\limits_{ - \infty }^z {n\left( {{z_1}} \right)} d{z_1}} \right) - \frac{{\left( {\nu  - 2} \right){{\left( {{{\tilde n}_{Pu}}} \right)}^2}}}{{1 - {{\tilde n}_{Pu}}}}\exp \left( { - \frac{1}{{v{{\tilde n}_{Pu}}}}\int\limits_{ - \infty }^z {n\left( {{z_1}} \right)d{z_1}} } \right).
\label{2.20}
\end{multline}  

So we got integro-differential equation. Let us introduce a new variable 
$N \left(z \right)$ into it, as follows:

\begin{equation}
N\left( z \right) \equiv \frac{1}{v}\int\limits_{ - \infty }^z {n\left( {{z_1}} \right)} d{z_1}.
\label{2.21}
\end{equation} 

Let us treat $N \left(z \right)$ as an analog of coordinate. Note that the
speed, i.e. a derivative of this coordinate, cannot be negative. In our case
the derivative is a neutron concentration, thus a negative sign at the speed
means the same sign at $n \left(z \right)$, which takes us out of the physical
region.

Taking into account (\ref{2.21}), (\ref{2.20}) becomes

\begin{multline}
\underbrace{\frac{{{d^2}N\left( z \right)}}{{d{z^2}}}}_{\text{resultant}} =\underbrace{ v\left( {1 - \frac{1}{\varepsilon }\exp \left( { - N\left( z \right)} \right)} \right)\frac{{dN\left( z \right)}}{{dz}}}_{\text{viscos force}} + \\
+\underbrace{ \left( {1 - \left( {\nu  - 2} \right){{\tilde n}_{Pu}}} \right)  + \frac{{(\nu -1) {{\tilde n}_{Pu}} - 1}}{{1 - {{\tilde n}_{Pu}}}}\exp \left( { - N\left( z \right)} \right) - \frac{{\left( {\nu  - 2} \right){{\left( {{{\tilde n}_{Pu}}} \right)}^2}}}{{1 - {{\tilde n}_{Pu}}}}\exp \left( { - \frac{{N\left( z \right)}}{{{{\tilde n}_{Pu}}}}} \right)}_{\text{potential force}}.
\label{2.22}
\end{multline} 

Let us note, that equation (\ref{2.22}) looks like this because we did not
neglect the derivative over time in the system of kinetic equations, when
switching to the autowave variable $z=y+vt$ in Section~\ref{section1}, as
it was done e.g. in~\citep{9,15,38,39,40,41}.

One could draw a parallel between equation (\ref{2.22}) and the Newton's
second law. If we consider the $N \left(z \right)$ an analog to coordinate, its
second derivative is the acceleration. So one has a resultant force scaled to
mass on the left in Eq.~(\ref{2.22}). On the right there is a sum of a viscous
force (a term including velocity) and a force caused by some potential of
interaction (which is the minus gradient of potential energy). It is easy to 
build an expression, the negative derivative of which would coincide with this
term. If we multiply (\ref{2.22}) by $\frac{{dN \left(z \right)}}{{dz}}$ and
transform it slightly, we get the following:

\begin{multline}
\frac{d}{{dz}}\left( {\frac{1}{2}{{\left( {\frac{{dN\left( z \right)}}{{dz}}} \right)}^2}} \right) = v\left( {1 - \frac{1}{\varepsilon }\exp \left( { - N\left( z \right)} \right)} \right){\left( {\frac{{dN\left( z \right)}}{{dz}}} \right)^2} + \\
 + \frac{d}{{dz}}\Bigg[ \left(1-\left(\nu-2\right)\tilde{n}_{Pu}\right)N\left(z\right) - \frac{(\nu -1)\tilde{n}_{Pu}-1}{1-\tilde{n}_{Pu}}\exp\left(-N\left(z\right)\right) +\\
+ \frac{\left(\nu-2\right)\left(\tilde{n}_{Pu}\right)^{3}}{1-\tilde{n}_{Pu}}\exp\left(-\frac{N\left(z\right)}{\tilde{n}_{Pu}}\right)\Bigg].
\label{2.23}
\end{multline}

Moving the derivative from the right side to the left, and grouping the
derivatives together, we obtain an equation similar to the energy conservation
law:

\begin{multline}
\frac{d}{{dz}}\Bigg[    {\frac{1}{2}{{\left( {\frac{{dN\left( z \right)}}{{dz}}} \right)}^2}}- \left(1-\left(\nu-2\right)\tilde{n}_{Pu}\right)N\left(z\right) +    \frac{(\nu -1)\tilde{n}_{Pu}-1}{1-\tilde{n}_{Pu}}\exp\left(-N\left(z\right)\right)+\\
+ \frac{\left(\nu-2\right)\left(\tilde{n}_{Pu}\right)^{3}}{1-\tilde{n}_{Pu}}\exp\left(-\frac{N\left(z\right)}{\tilde{n}_{Pu}}\right)\Bigg] =v\left( {1 - \frac{1}{\varepsilon }\exp \left( { - N\left( z \right)} \right)} \right){\left( {\frac{{dN\left( z \right)}}{{dz}}} \right)^2}.
\label{2.24}
\end{multline}

A derivative of the sum of kinetic and potential energy in the
expression~(\ref{2.24}) equals to the viscous force. Thus this equation may be
considered as the energy conservation law, and the total energy may change only
through the work of viscous force. Let us introduce the notation for the 
potential energy and the terms it contains:

\begin{eqnarray}
{k_1} \equiv  - \left( {1 - \left( {\nu  - 2} \right){{\tilde n}_{Pu}}} \right),{k_2} \equiv \frac{{(\nu -1) {{\tilde n}_{Pu}} - 1}}{{1 - {{\tilde n}_{Pu}}}},{k_3} \equiv  - \frac{{\left( {\nu  - 2} \right){{\left( {{{\tilde n}_{Pu}}} \right)}^3}}}{{1 - {{\tilde n}_{Pu}}}}\label{2.25}\\
U\left( N \right) \equiv {k_1}N + {k_2}\exp \left(- N\right)-{k_3}\exp \left( - \frac{N}{\tilde {n}_{Pu}} \right).\label{2.26}
\end{eqnarray}

So, we obtain an explicit expression of the energy conservation law.

\begin{equation}
\frac{d}{{dz}}\left( \frac{1}{2}{{\left( {\frac{{dN\left( z \right)}}{{dz}}} \right)}^2} + U\left( N \right) \right) = v\left( {1 - \frac{1}{\varepsilon }\exp \left( { - N\left( z \right)} \right)} \right){\left( {\frac{{dN\left( z \right)}}{{dz}}} \right)^2}.
\label{2.27}
\end{equation}

We are interested in finding the cases of equilibrium, which corresponds to the
autowave fission mode. This happens when the total force consisting of the 
"potential" force and "dissipative" viscous force, is zero. For this the 
potential energy must have a point of minimum, because the derivative (the
potential force) is zero at this point. The speed (and hence the kinetic
energy) must also be zero at this minimum point, meaning the dissipative
viscous force is zero.

Let us take a look at the behavior of the potential energy~(\ref{2.26})
depending on its coefficients.
 
First, we consider the case when ${\tilde n}_{Pu}$ is less than one. Please
note that in contrast to~\citep{9}, where the value of ${\tilde n}_{Pu}$ is
considered to be the equilibrium concentration of plutonium, and therefore
cannot be greater than unity, in the present paper we consider 
${\tilde n}_{Pu}$ to be an arbitrary parameter, so it may be smaller or greater
than one.

In this case (when $ {\tilde n}_{Pu}$ is smaller than one) the exponent with
${\tilde n}_{Pu}$ in its index makes a significantly smaller contribution to
the potential energy~(\ref{2.26}) than another one does. Therefore $k_3$ may be
neglected, and the largest contribution is made by $k_1$ and $k_2$.

\begin{figure}[tb!]
\begin{center}
\includegraphics[width=1.0\linewidth]{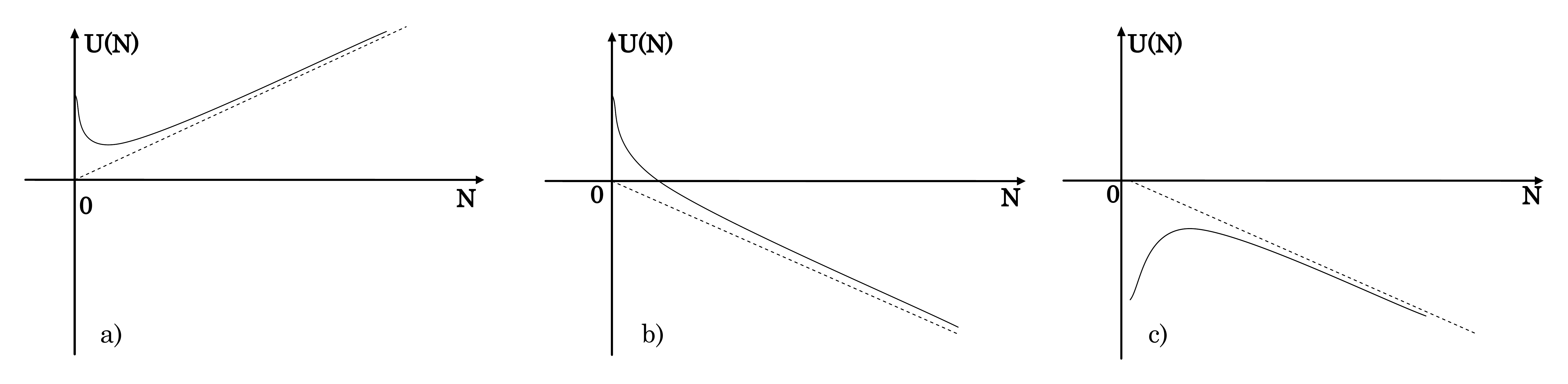}
\end{center}
\caption{A sketch of various analogs of potential energy, provided that 
${\tilde n}_{Pu}<1$. a) $k_1>0$, $k_2>0$; b) $k_1<0$, $k_2>0$; c) $k_1<0$, 
$k_2<0$}
\label{fig2.1}
\end{figure}

Fig.~\ref{fig2.1}(a) shows the form of potential energy when 
$\left(k_1> 0, k_2> 0 \right)$. There is an apparent point of minimum in this
graph -- a stable stationary point. The derivative of potential energy (the potential
force) is obviously zero at this point. So this is a possible stationary state,
if the kinetic energy is zero, and corresponds to the autowave fission mode.

\begin{figure}[tb!]
\begin{center}
\includegraphics[width=0.8\linewidth]{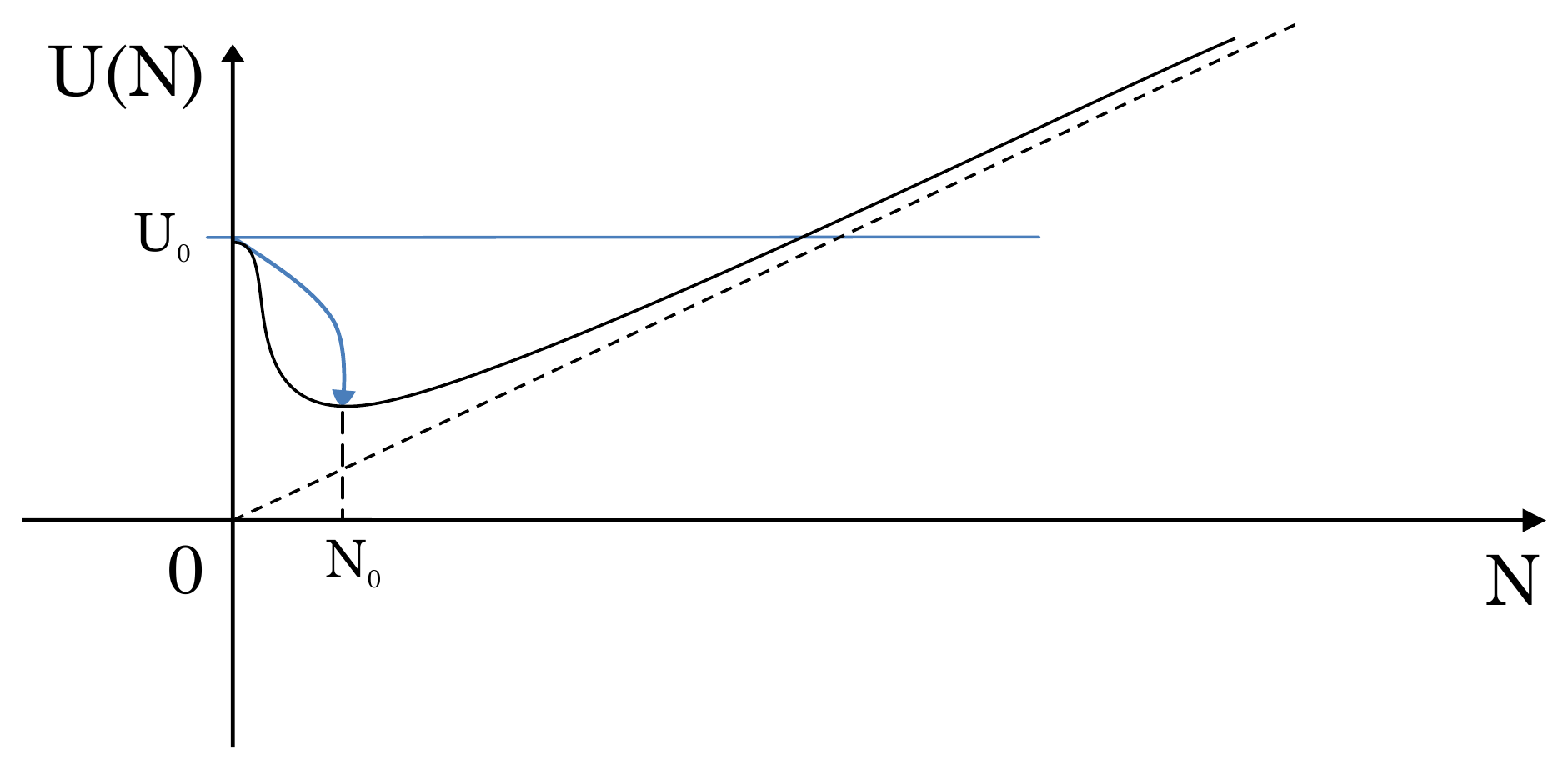}
\end{center}
\caption{A single possible form of potential energy that allows the existence
of an auto-wave mode is shown in black. The arrow shows the path of total
energy along which this mode can be set.}
\label{fig2.2}
\end{figure}

Remind that the derivative of $N$ cannot be negative, because it would lead us
to negative neutron concentration, which is non-physical. If we treat $N$ as an
analog of some coordinate, then its derivative is the speed. So this speed
cannot be negative -- as non-physical. If the object starts from the point 
$N=0$ and we want it to stop at the point of equilibrium (minimum), it should
not overshoot this point. Otherwise the derivative of the potential energy will
be non-zero, the force will act on the body, and the body will keep moving. It
is known that the potential energy cannot be greater than total energy, so at
some moment, when the kinetic energy completely transforms into potential, the
object will start moving in the opposite direction. This means that the speed
of the body will become negative. And we cannot allow this as non-physical.
Therefore it is necessary for the entire kinetic energy to dissipate due to
viscosity by the time when the potential energy reaches a point of minimum.
This way point the object stops at a stationary point. Graphically this path is
shown in Fig.~\ref{fig2.2}.

Next, consider the case when $\left(k_1 <0, k_2> 0 \right)$
(Fig.~\ref{fig2.1}(b)). There is no point of minimum in this case, and the
potential energy tends to $\left(- \infty \right)$. It means that this system
has no equilibrium point, and the autowave mode of nuclear burning is
impossible. So this case is not an option for us.

And the last case is $\left(k_1 <0, k_2 <0 \right)$, which is shown in
Fig.~\ref{fig2.1}(c). There is one stationary point in such system -- a point
of maximum. However, this point is unstable, and the slightest deviation from
this point is able to direct the system either towards zero, or to 
$\left(- \infty \right)$. So in this case the system cannot reach the
equilibrium state and stay in it, which means the autowave mode is also not
possible.

At this point we can conclude that the only case suitable for our purpose, when
${\tilde n}_{Pu}<1$, is the one shown in Fig.~\ref{fig2.1}(a), i.e. when 
$\left(k_1> 0, k_2>0 \right)$.

Let us now consider the situation when the parameter $\tilde{n}_{Pu}$ is
greater than one. In this case $k_1$ and $k_3$ are always greater than zero,
and $k_2$ is always less than zero. The behavior of the potential energy near
the zero point will depend mostly on the exponent with $k_3$ factor. This
factor is certainly greater than zero when $\tilde{n}_{Pu}>1$, so one would
expect a point of minimum on the graph, which is exactly what we need. The 
behavior of $U\left(N \right)$ will look similar to what is shown in
Fig.~\ref{fig2.1}(a). All the arguments related to Fig.~\ref{fig2.2} are also
true for this case. So any case, when $\tilde{n}_{Pu}$ is greater than one,
will do.

\section{Finding a criterion for the wave velocity}

\begin{figure}[tb!]
\begin{center}
\includegraphics[width=0.6\linewidth]{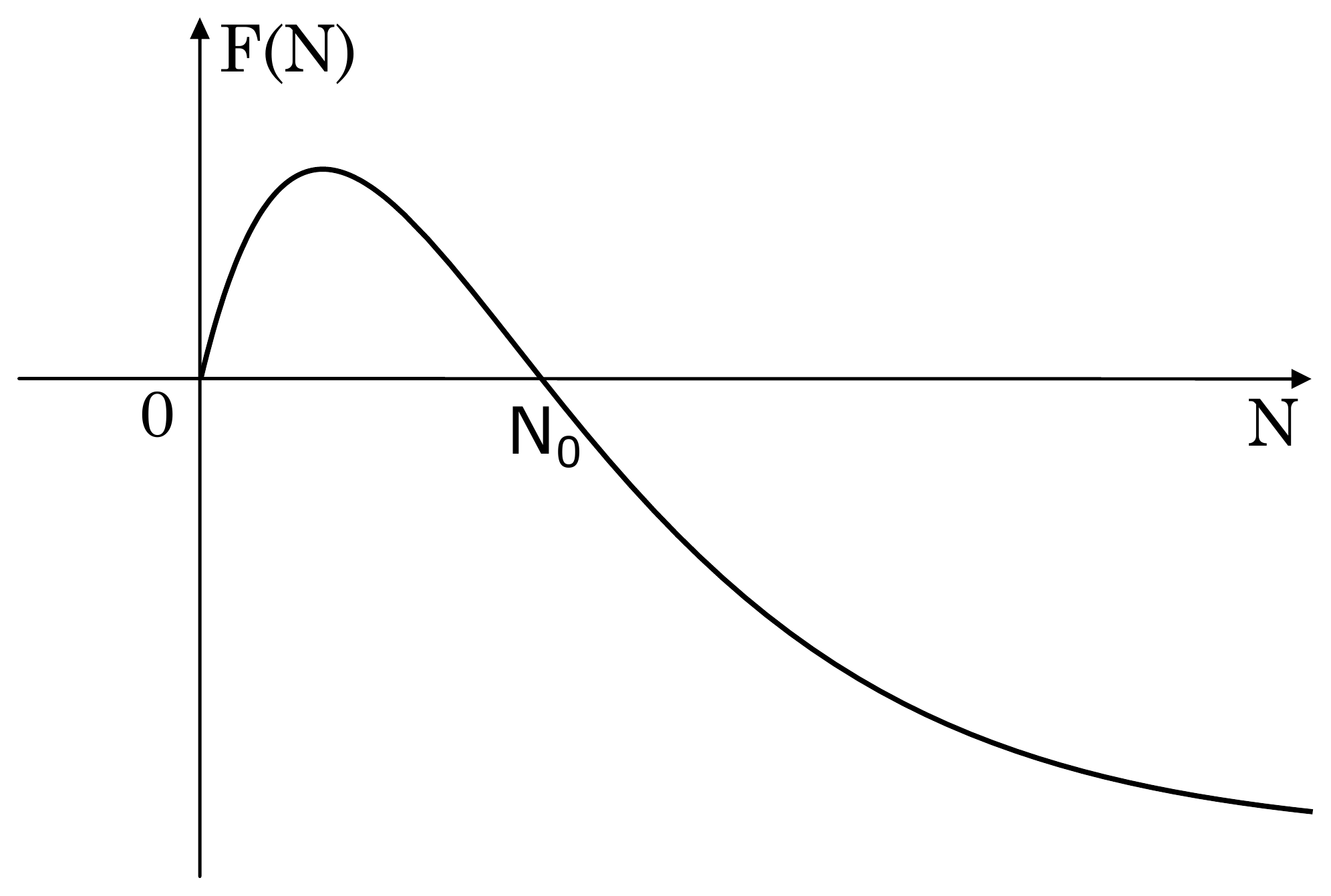}
\end{center}
\caption{A schematic representation of the potential force $F \left(N \right)$
(a derivative of $U(N)$ taked with the opposite sign), when the potential
energy has a minimum point.}
\label{fig2.3}
\end{figure}

We determined that in order for the autowave mode to be possible, the
coefficients $k_1, k_2$ and $k_3$ (taking into account the value 
$\tilde{n}_{Pu}$) must be chosen in such a way, that the potential energy
(or its analog) has a point of minimum. The potential force (minus derivative
of $U(N)$) in this case will look like Fig.~\ref{fig2.3}.

\begin{figure}[tb!]
\begin{center}
\includegraphics[width=0.6\linewidth]{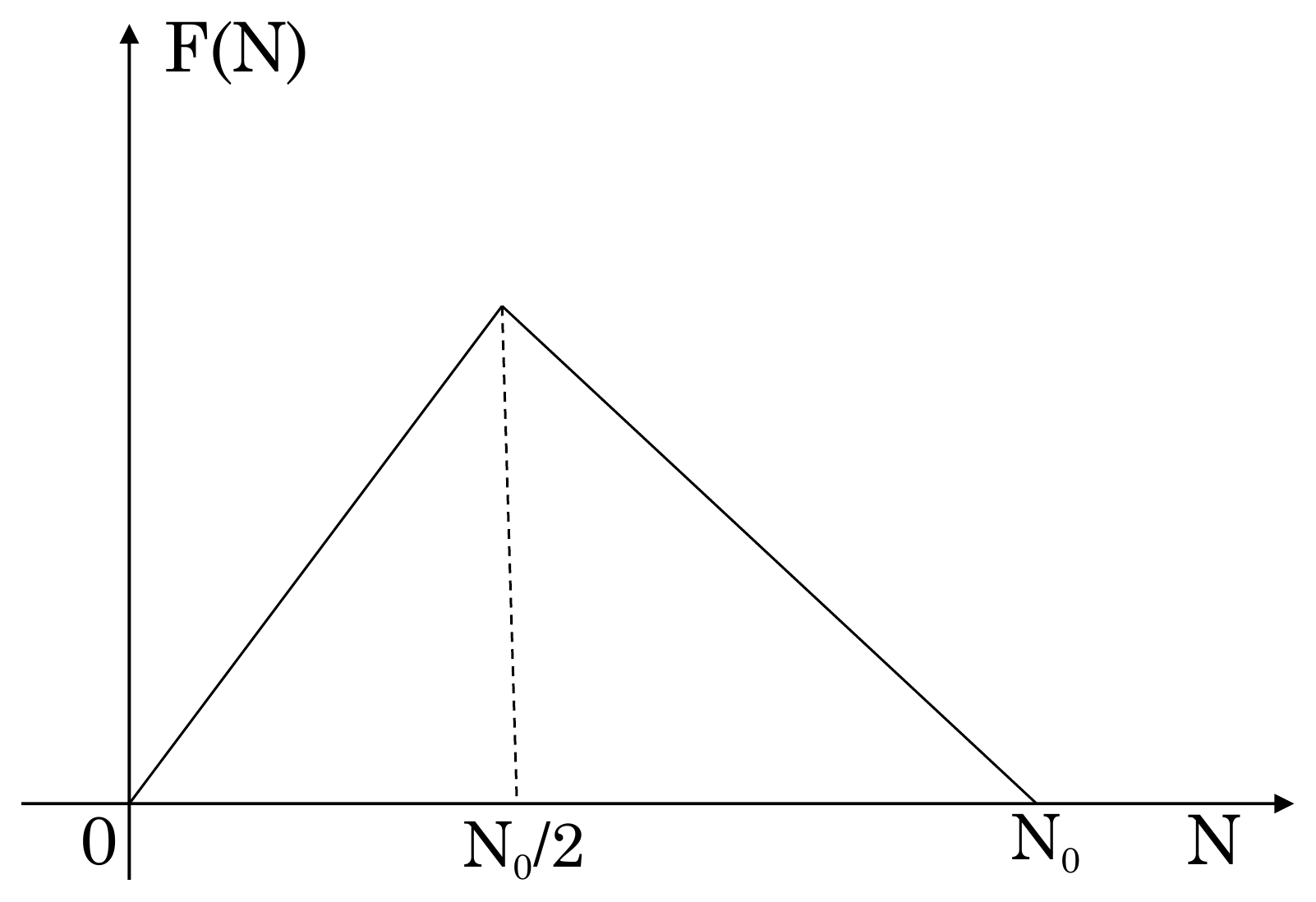}
\end{center}
\caption{Approximate $F\left(N\right)$ which substitutes the real one in
Fig.~\ref{fig2.3}.}
\label{fig2.4}
\end{figure}

Since in the autowave form the wave propagates over an infinite interval, and
there is no explicit dependence on variable $z$ in equation~(\ref{2.27}), it is
always possible to perform a variable substitution that shifts $z$ relative to
the function $N$. Let us thus switch to a new variable so that $z=0$ at the
point $\frac{N_0}{2}$. We replace the function of force by a triangular one
(Fig.~\ref{fig2.4}).

Next, we try to replace the function of force with two straight lines -- one on
the negative part of the coordinate axis $z$, and the other on the positive
part. If we solve the equation~(\ref{2.27}) with respect to $N$ with the
functions of force in the form of two lines, we obtain two unknown constants
for each one of solutions (because the equations are of second order). We
adjust these constants to satisfy the boundary conditions. We also hope that
the constants adjustment will end up with zeroing out one of them for each
solution, since the functions at which they appear will diverge to infinity.
Then, if we join the functions and their first derivatives at the point 
$\frac{N_0}{2}$, we would obtain one spare equation, which can be used to find
the wave speed.

So, let us first consider the interval $z\in\left(-\infty,0\right)$. Here we
make the following substitution:

\begin{eqnarray}
F\left( N \right) = qN,\label{2.28}\\
U\left( N \right) = U_0 - q\frac{{{N^2}}}{2}, 
\label{2.29}
\end{eqnarray}

\noindent
where $q$ is the slope, and is equal to:

\begin{equation}
q = -{\left. {\frac{{{d^2}U\left( N \right)}}{{d{N^2}}}} \right|_{N = 0}} =  -\left({{k_2} + \frac{{{k_3}}}{{{{\tilde n}_{Pu}}^2}}} \right)=1.
\label{2.30}
\end{equation}

By substituting~(\ref{2.29}) into~(\ref{2.27}) we obtain a homogeneous
differential equation of second order, whose solution looks like:

\begin{eqnarray}
N\left( z \right) = {C_1}\exp \left( {{\chi _1}z} \right) + {C_2}\exp \left( {{\chi _2}z} \right)\label{2.31}\\
{\chi _{1,2}} = \frac{1}{2}\left( { - \frac{{v{n_{8, - \infty }}}}{\varepsilon } \pm \sqrt {{{\left( {\frac{{v{n_{8, - \infty }}}}{\varepsilon }} \right)}^2} + 4q} } \right).
\label{2.32}
\end{eqnarray}

$\chi_1$ is apparently positive, and $\chi_2$ is negative. So in order for the
$N\left(z\right)$ to converge to $( -\infty )$, $C_2$ must be zero. So

\begin{equation}
N\left( z \right) = {C_1}\exp \left( {{\chi _1}z} \right).
\label{2.33}
\end{equation}

Now let us consider the interval $z\in\left(0,+\infty\right)$. In this case
we make the following substitution:

\begin{eqnarray}
F\left( N \right) =  - \left( {N - {N_0}} \right),\label{2.34}\\
U\left( N \right) = U\left( {{N_0}} \right) - \frac{1}{2}{\left( {N - {N_0}} \right)^2}.
\label{2.35}
\end{eqnarray}

If we substitute the potential energy again into~(\ref{2.27}), we obtain a
similar solution.

\begin{eqnarray}
N\left( z \right) = {N_0} + {D_1}\left( v \right)\exp \left( {{\lambda _1}\left( v \right)z} \right) + {D_2}\left( v \right)\exp \left( {{\lambda _2}\left( v \right)z} \right),\label{2.36}\\
{\lambda _{1,2}} = \frac{1}{2}\left( { - v\frac{{{n_{8, - \infty }}}}{\varepsilon }\exp \left( { - {N_0}} \right) \pm \sqrt {{{\left( {\frac{{v{n_{8, - \infty }}}}{\varepsilon }\exp \left( { - {N_0}} \right)} \right)}^2} - 4} } \right).
\label{2.37}
\end{eqnarray}

As we can see, both $\lambda_1$ and $\lambda_2$ are negative, so at $(+\infty)$
both constants are at the "good" exponents (the converging ones), and we cannot
zero out any of them. So this way we do not obtain an additional equation,
which could be used to find the wave speed.

There is an expression that may be negative under the root in
equation~(\ref{2.37}). Thus, the solution will be the sum of sine and cosine
with some coefficients. Since they have a finite period, and the wave
propagates on an infinite interval, the derivative will become negative at some
point. As we noted above, this would lead us to the negative neutron
concentration, which is non-physical. So we require that

\begin{equation}
{\left( {\frac{{v}}{\varepsilon }\exp \left( { - {N_0}} \right)} \right)^2} - 4 \ge 0.
\label{2.38}
\end{equation}

Considering this requirement relative to $v$, we obtain the following
restriction:

\begin{equation}
v \ge 2\varepsilon \exp \left( {{N_0}} \right).
\label{2.39}
\end{equation}

It means, that the wave velocity may take a continuous spectrum of values
greater than certain minimum value. In the framework of the analogy to the
energy conservation law, one might say that the viscosity cannot be less
than a certain value. It makes sense, because with higher viscosity the
kinetic energy will still dissipate completely to the minimum point, but if the
viscosity is not high enough, the body will possess some nonzero speed at the
stationary point, which eventually leads to the reverse motion. And as we noted
above, the reverse motion is not allowed as non-physical.

\begin{figure}[tb!]
\begin{center}
\includegraphics[width=0.6\linewidth]{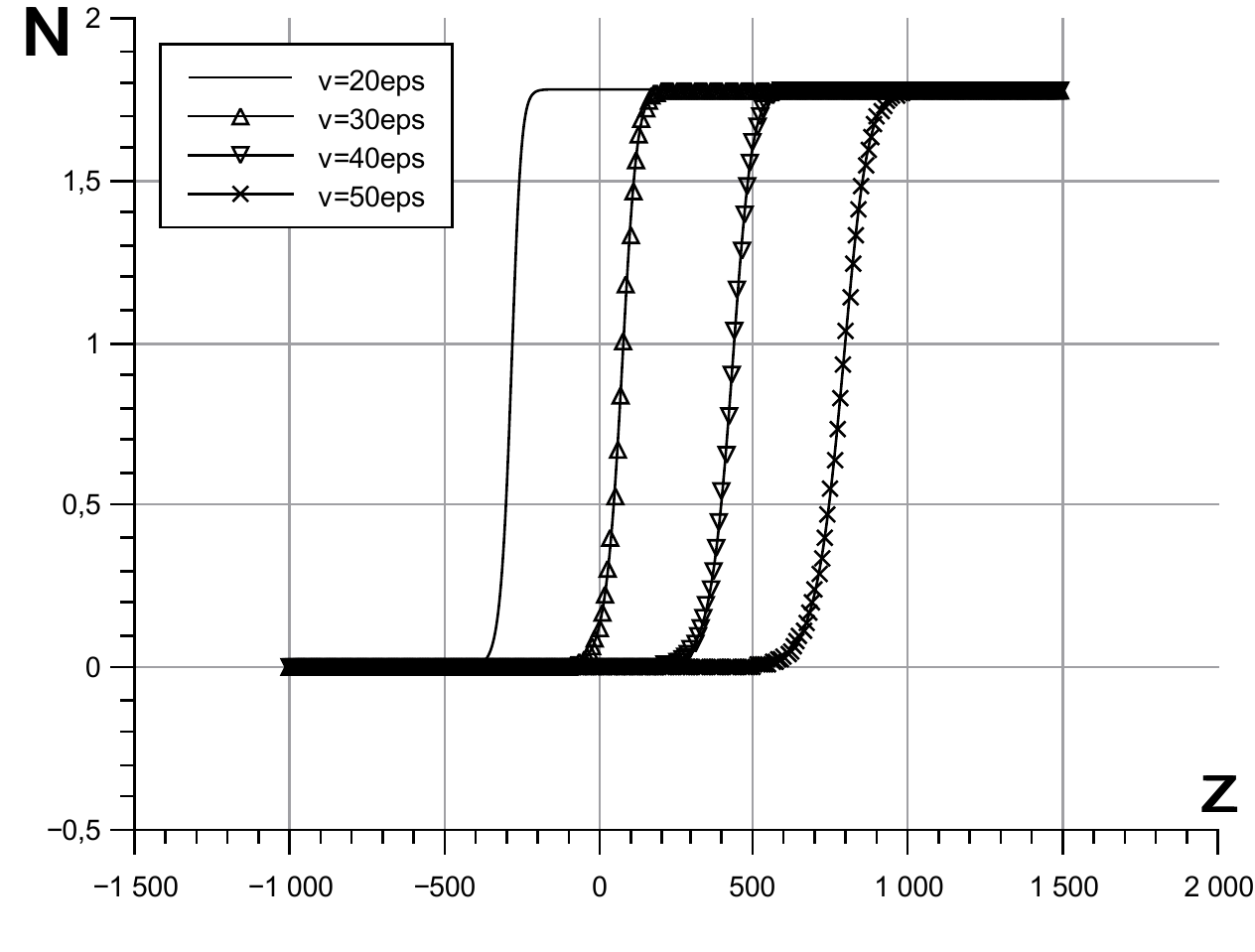}
\end{center}
\caption{Solutions of Eq.~(\ref{2.22}) as a function $N(z)$ when a minimum
exists in potential energy. The curves are given for different velocities that
satisfy the requirement~(\ref{2.39}).}
\label{figN}
\end{figure}

Let us check if the speed can take different values. To do this, we choose the
parameters so that the potential energy has the point of minimum, and solve the
equation~(\ref{2.22}) numerically. These parameters are: 
${{\tilde n} _{Pu}}=0.74, \nu=2.8, N_0=1.8$, where $N_0$ is the point of
potential energy minimum. With such values, the condition~(\ref{2.39}) has the
form: $v \ge 15\varepsilon$. Fig.~\ref{figN} shows the dependences of $N$ on
$z$ for different velocities. Comparing it to the potential energy 
(Fig.~\ref{2.1}(a)), one cant notice the following: the coordinate is initially
zero, and then it goes to the point of the potential energy minimum. The change
in speed, i.e. dissipative force, affects the form of this transition and
shifts it in time.

\section{Finding the spectrum of neutron energies, suitable for an auto-wave mode}

In order for the autowave to exist, the potential
energy (or its analog) must have a point of minimum, and also the wave speed
must be such that the kinetic energy is completely dissipated, when the
potential energy reaches this minimum point. So now as we realize that, it is
interesting to study how the fulfillment of the first condition (existence of
the minimum point) depends on the neutrons energy. According to~(\ref{2.25})
and~(\ref{2.26}), $U\left(N\right)$ depends on $\nu$ and $\tilde{n}_{Pu}$.
These parameters, in turn, depend on the neutron energy. We find the parameter
$\tilde{n}_{Pu}$ using Eq.~(\ref{9.3.3}) and examine it depending on the
neutron energies.


According to~\citep{44,45,46}, the mean number of instantaneous neutrons $\nu$
produced by single fission has the following dependence on energy:

\begin{multline}
\nu \left( {Z,A,{E_n}} \right) = 2.33 + 0.06\left( {2 - {{\left( { - 1} \right)}^{A + 1 - Z}} - {{\left( { - 1} \right)}^Z}} \right) + 0.15\left( {Z - 92} \right)+\\
 + 0.02\left( {A - 235} \right) + \left( {0.13 + 0.006\left( {A - 235} \right)} \right)\left( {{E_n} - {E_{threshold}}} \right).
\label{2.41}
\end{multline}

Since we are interested in the average number of neutrons $\nu$ for
$ Pu^{239}$, we choose the following parameters for the equation~(\ref{2.41}):

\begin{equation}
A = 239, Z = 94,E_{threshold}\ = -0.89\left(MeV\right).
\label{2.42}
\end{equation}


So now we know the dependence of the coefficients in the potential energy on
the energy of the neutrons. We have to find the entire spectrum of neutron 
energies at which the potential energy has a point of minimum. It may be done
as follows. Break some interval of $N$ (from zero up to a certain maximum
value) into sufficiently small segments, and compare the values of 
$U\left(N\right)$ at three adjacent points. If the value at the middle point
is less than those at both ends, then the minimum of the function exists. If we
do not find such three points, then there is no minimum at this energy. Having
analyzed the entire spectrum of neutron energies this way, we find all the
energy intervals with minima.

We show all the neutron energies, for which we found the minimum of potential
energy, as a set of points in the graph, where the neutron energy is along the
X-axis, and the parameter $\tilde{n}_{Pu}$ is along the Y-axis. Only the
values, for which there is a stationary point in the potential energy, are
shown.

\begin{figure}[tb!]
\begin{center}
\includegraphics[width=1.0\linewidth]{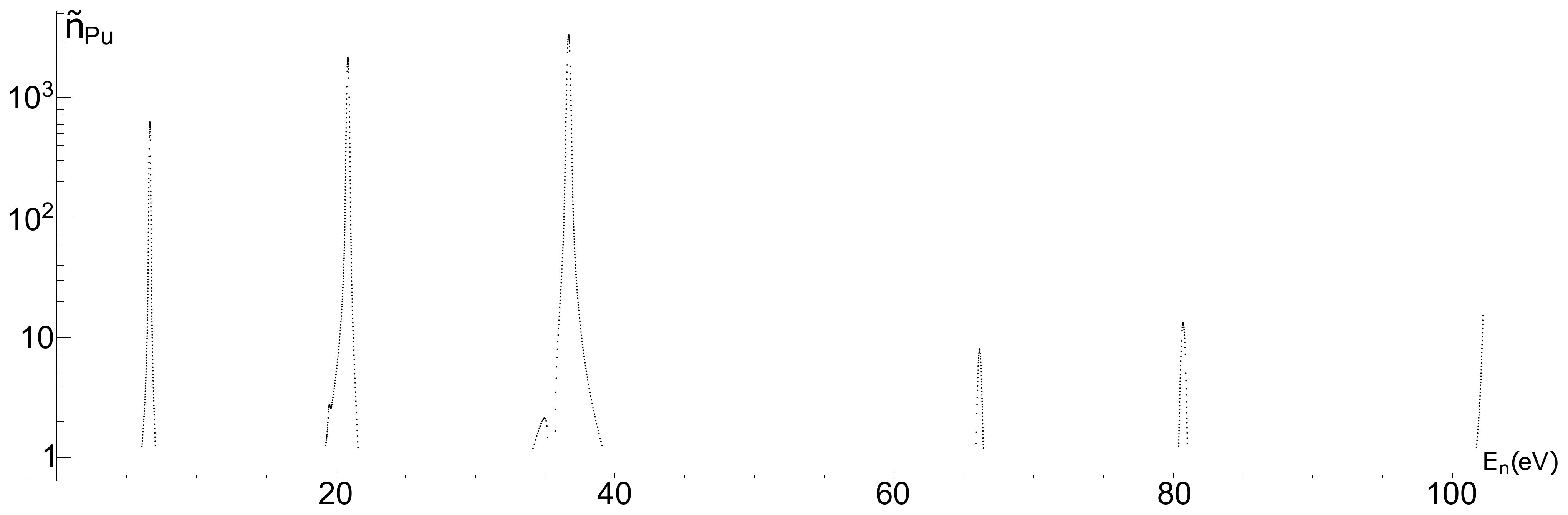}
\end{center}
\caption{The dependence of the parameter $\tilde{n}_{Pu}$ on the energy of
neutrons, for the energies, when the potential energy has the point of minimum,
and the autowave may exist. The image shows the energy range from 0 to 100~eV.}
\label{fig2.5}
\end{figure}

Fig.~\ref{fig2.5} shows the values of the parameter $\tilde{n}_{Pu}$ for
neutron energies in the range from 0 to 100~eV. We used the dependence of the
cross-sections of nuclides on the neutron energy from the database~\citep{47}.
As seen from Fig.~\ref{fig2.5} (according to the considerations in
Section~\ref{section2}), there are seven regions at about 6-7~eV, 19-21~eV,
34-40~eV, 66-67~eV, 81-82~eV and 90-100~eV, having a minimum of potential
energy, which makes the autowave of nuclear fission possible.

\begin{figure}[tb!]
\begin{center}
\includegraphics[width=1.0\linewidth]{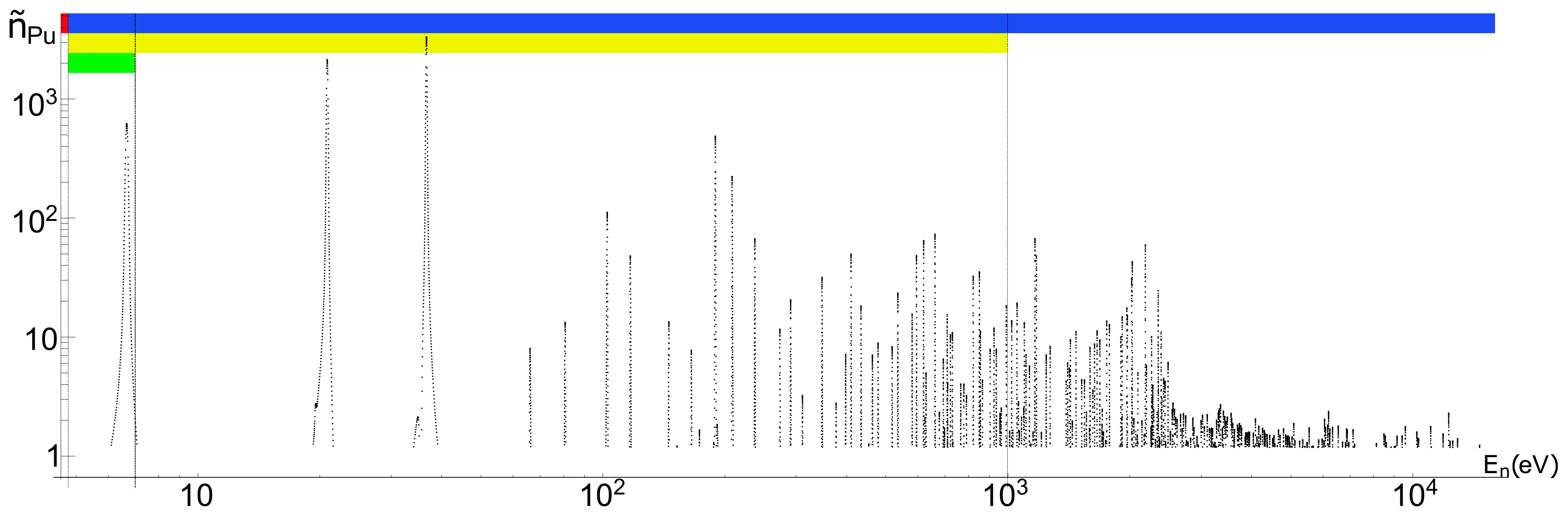}
\end{center}
\caption{The dependence of the parameter $\tilde{n}_{Pu}$ on the energy of
neutrons, for the energies, when the potential energy has the point of minimum,
and the existence of an autowave is possible. The upper part of the image marks
different neutron energy regions with different colors. \textbf{Red} is for
thermal neutrons (there are no points in this area here, so it is barely
noticeable), \textbf{blue} is for intermediate neutrons (this area goes beyond
the graph and therefore is not depicted fully). There are no points in the
region of fast neutrons, so it is not shown here. The additional markings are
for resonance neutrons (\textbf{yellow}) and epithermal neutrons
(\textbf{green}). This picture embraces all possible points, as the entire
region of neutron energies from experimental data~\citep{47} was studied}
\label{fig2.6}
\end{figure}

The entire spectrum of neutron energies is shown in Fig.~\ref{fig2.6}.

\section{Numerical simulation}

The system of equations studied above is simplified, and does not fully reflect
the physical processes that take place in the reactor core. Let us consider the
system of equations given in~\citep{32}, which is more accurate (though not
ideal), and consists of 18 equations. The kinetic equation for the neutron
concentration

\begin{equation}
\frac{{\partial \,n\,\left( {x,t} \right)}}{{\partial \,t}} = D\,\frac{{{\partial ^2}\,n\,\left( {x,t} \right)}}{{\partial {x^2}}} + q\left( {x,t} \right),
\label{5.1}
\end{equation}

\noindent
where $q(x,t)$ is the internal source of neutrons, which has the form:

\begin{multline}
q\left( {x,t} \right) = \left[ {{\nu ^{\left( {Pu} \right)}}\left( {1 - {p^{\left( {Pu} \right)}}} \right) - 1} \right] \cdot n\left( {x,t} \right) \cdot {V_n} \cdot \sigma _f^{\left( {Pu} \right)} \cdot {N_{Pu}}(x,t) + \\
 + \left[ {{\nu ^{\left( 5 \right)}}\left( {1 - {p^{\left( 5 \right)}}} \right) - 1} \right] \cdot n(x,t) \cdot {V_n} \cdot \sigma _f^{\left( 5 \right)} \cdot {N_5}(x,t) + \\
 + \ln 2 \cdot \sum\limits_{i = 1}^6 {\left[ {\frac{{\tilde N_i^{\left( {Pu} \right)}(x,t)}}{{T_i^{\left( {Pu} \right)}}} + \frac{{\tilde N_i^{\left( 5 \right)}(x,t)}}{{T_i^{\left( 5 \right)}}}} \right]}  - n(x,t) \cdot {V_n} \cdot \sum\limits_{5,8,9,Pu} {\sigma _c^{\left( i \right)} \cdot {N_i}(x,t)}  - \\
 - n(x,t) \cdot {V_n} \cdot \sum\limits_{i = 1}^6 {\left[ {\sigma _c^{i\left( {Pu} \right)} \cdot \tilde N_i^{\left( {Pu} \right)}(x,t) + \sigma _c^{i\left( 5 \right)} \cdot \tilde N_i^{\left( 5 \right)}(x,t)} \right]}  - \\
 - n(x,t) \cdot {V_n} \cdot \sigma _c^{eff\left( {Pu} \right)} \cdot {{\bar N}^{\left( {Pu} \right)}}(x,t) - n(x,t) \cdot {V_n} \cdot \sigma _c^{eff\left( 5 \right)} \cdot {{\bar N}^{\left( 5 \right)}}(x,t).
\label{5.2}
\end{multline}

In equations~(\ref{5.1}) and~(\ref{5.2}) $n(x,t)$ is the neutron concentration;
$V_n$ is the neutron velocity; $\nu^{(Pu)}$ and $\nu^{(5)}$ are the mean number
of instantaneous neutrons per $Pu^{239}$ and $^{235}U$ fission respectively.
$N_5$, $N_8$, $N_9$ and $N_{Pu}$ are the $^{235}U$, $^{238}U$, $^{239}U$ and
$^{235}Pu$ concentrations respectively;
$\tilde{N}_{i}^{Pu}$ and $\tilde{N}_{i}^{5}$ are the concentrations of
fragments with neutron excess, produced due to $^{239}Pu$ and $^{235}U$ fission
respectively;
$\overline{N}^{(Pu)}$ and $\overline{N}^{(5)}$ are the concentrations of all
the other fission fragments of $^{239}Pu$ and $^{235}U$ respectively;
$\sigma_c$ and $\sigma_f $ are the microscopic cross-sections of the neutron
capture and nuclear fission;
the parameters $p_i$ $\left(p=\sum\limits_{i = 1}^6p_i\right)$ and $T_{1/2}$
characterize the groups of delayed neutrons. They are well known and presented
in~\citep{46}.
$\sigma_c^{eff}$ is some effective microscopic cross-section of neutron capture
by fragments.

Let $N_{8,0}$ be the initial $^{238}U$ concentration. Then we introduce the
new dimensionless concentrations, equal to the old ones divided by $N_{8,0}$.

Then~(\ref{5.1}), taking into account~(\ref{5.2}) and using the dimensionless
concentrations, becomes:

\begin{multline}
\frac{{\partial \,n\,\left( {x,t} \right)}}{{\partial \,t}} = D\,\frac{{{\partial ^2}\,n\,\left( {x,t} \right)}}{{\partial {x^2}}} + {V_n}\sigma _f^{\left( {Pu} \right)}\left[ {{\nu ^{\left( {Pu} \right)}}\left( {1 - {p^{\left( {Pu} \right)}}} \right) - 1} \right]{N_{8,0}}n\left( {x,t} \right){N_{Pu}}(x,t) + \\
 + \left[ {{\nu ^{\left( 5 \right)}}\left( {1 - {p^{\left( 5 \right)}}} \right) - 1} \right]{N_{8,0}}{V_n}\sigma _f^{\left( 5 \right)}n(x,t){N_5}(x,t) + \\
 + \ln 2\sum\limits_{i = 1}^6 {\left[ {\frac{{\tilde N_i^{\left( {Pu} \right)}(x,t)}}{{T_i^{\left( {Pu} \right)}}} + \frac{{\tilde N_i^{\left( 5 \right)}(x,t)}}{{T_i^{\left( 5 \right)}}}} \right]}  - {N_{8,0}}{V_n}n(x,t)\sum\limits_{5,8,9,Pu} {\sigma _c^{\left( i \right)}{N_i}(x,t)}  - \\
 - {N_{8,0}}{V_n}n(x,t)\sum\limits_{i = 1}^6 {\left[ {\sigma _c^{i\left( {Pu} \right)}\tilde N_i^{\left( {Pu} \right)}(x,t) + \sigma _c^{i\left( 5 \right)}\tilde N_i^{\left( 5 \right)}(x,t)} \right]}  - \\
 - {N_{8,0}}{V_n}\sigma _c^{eff\left( {Pu} \right)}n(x,t){{\bar N}^{\left( {Pu} \right)}}(x,t) - {N_{8,0}}{V_n}\sigma _c^{eff\left( 5 \right)}n(x,t){{\bar N}^{\left( 5 \right)}}(x,t).
\label{5.3}
\end{multline}

Then let us switch to dimensionless coordinates. Since

\begin{equation}
V_n \sigma_j^{(i)}N_{8,0}=\left[\frac{cm}{s}\cdot cm^2\cdot\frac{1}{cm^3}\right]=\left[\frac{1}{s}\right],
\label{5.4}
\end{equation}

\noindent
the dimensionless time may be introduced as follows:

\begin{equation}
t_{old}=\frac{1}{V_n \sigma_c^{(8)}N_{8,0}}t_{new}.
\label{5.5}
\end{equation}

Given (\ref{5.5}), the equation (\ref{5.3}) takes the form:

\begin{multline}
\frac{{\partial \,n\,\left( {x,t} \right)}}{{\partial \,t}} = \,\frac{D}{{{V_n}\sigma _c^{\left( 8 \right)}{N_{8,0}}}}\frac{{{\partial ^2}\,n\,\left( {x,t} \right)}}{{\partial {x^2}}} + \frac{{\sigma _f^{\left( {Pu} \right)}}}{{\sigma _c^{\left( 8 \right)}}}\left[ {{\nu ^{\left( {Pu} \right)}}\left( {1 - {p^{\left( {Pu} \right)}}} \right) - 1} \right]n\left( {x,t} \right){N_{Pu}}(x,t) + \\
 + \frac{{\sigma _f^{\left( 5 \right)}}}{{\sigma _c^{\left( 8 \right)}}}\left[ {{\nu ^{\left( 5 \right)}}\left( {1 - {p^{\left( 5 \right)}}} \right) - 1} \right]n(x,t){N_5}(x,t) +\\
+ \frac{1}{{{V_n}\sigma _c^{\left( 8 \right)}{N_{8,0}}}}\ln 2\sum\limits_{i = 1}^6 {\left[ {\frac{{\tilde N_i^{\left( {Pu} \right)}(x,t)}}{{T_i^{\left( {Pu} \right)}}} + \frac{{\tilde N_i^{\left( 5 \right)}(x,t)}}{{T_i^{\left( 5 \right)}}}} \right]}  - \\
 - n(x,t)\sum\limits_{5,8,9,Pu} {\frac{{\sigma _c^{\left( i \right)}}}{{\sigma _c^{\left( 8 \right)}}}{N_i}(x,t)}  - n(x,t)\sum\limits_{i = 1}^6 {\left[ {\frac{{\sigma _c^{i\left( {Pu} \right)}}}{{\sigma _c^{\left( 8 \right)}}}\tilde N_i^{\left( {Pu} \right)}(x,t) + \frac{{\sigma _c^{i\left( 5 \right)}}}{{\sigma _c^{\left( 8 \right)}}}\tilde N_i^{\left( 5 \right)}(x,t)} \right]}  - \\
 - \frac{{\sigma _c^{eff\left( {Pu} \right)}}}{{\sigma _c^{\left( 8 \right)}}}n(x,t{{\bar N}^{\left( {Pu} \right)}}(x,t) - \frac{{\sigma _c^{eff\left( 5 \right)}}}{{\sigma _c^{\left( 8 \right)}}}n(x,t){{\bar N}^{\left( 5 \right)}}(x,t).
\label{5.6}
\end{multline}

We introduce the dimensionless coordinate $x$ as follows:

\begin{equation}
x_{old}=\sqrt{\frac{D}{V_n \sigma_c^{(8)}N_{8,0}}}x_{new}.
\label{5.7}
\end{equation}

Designating

\begin{equation}
\varepsilon _i^{\left( j \right)} \equiv \frac{{\sigma _i^{\left( j \right)}}}{{\sigma _c^{\left( 8 \right)}}},\beta _i^{\left( j \right)} \equiv \frac{{\ln 2}}{{{V_n}\sigma _c^{\left( 8 \right)}{N_{8,0}}T_i^{\left( j \right)}}}.
\label{5.8}
\end{equation}

\noindent
and taking into account~(\ref{5.7}) and~(\ref{5.8}), we transform
equation~(\ref{5.6}) to the form:

\begin{multline}
\frac{{\partial \,n\,\left( {x,t} \right)}}{{\partial \,t}} = \,\frac{{{\partial ^2}\,n\,\left( {x,t} \right)}}{{\partial {x^2}}} + \varepsilon _f^{\left( {Pu} \right)}\left[ {{\nu ^{\left( {Pu} \right)}}\left( {1 - {p^{\left( {Pu} \right)}}} \right) - 1} \right]n\left( {x,t} \right){N_{Pu}}(x,t) + \\
 + \varepsilon _f^{\left( 5 \right)}\left[ {{\nu ^{\left( 5 \right)}}\left( {1 - {p^{\left( 5 \right)}}} \right) - 1} \right]n(x,t){N_5}(x,t) + \sum\limits_{i = 1}^6 {\left[ {\beta _i^{\left( {Pu} \right)}\tilde N_i^{\left( {Pu} \right)}(x,t) + \beta _i^{\left( 5 \right)}\tilde N_i^{\left( 5 \right)}(x,t)} \right]}  - \\
 - n(x,t)\sum\limits_{5,8,9,Pu} {\varepsilon _c^{\left( i \right)}{N_i}(x,t)}  - n(x,t)\sum\limits_{i = 1}^6 {\left[ {\varepsilon _c^{i\left( {Pu} \right)}\tilde N_i^{\left( {Pu} \right)}(x,t) + \varepsilon _c^{i\left( 5 \right)}\tilde N_i^{\left( 5 \right)}(x,t)} \right]}  - \\
 - \varepsilon _c^{eff\left( {Pu} \right)}n(x,t){{\bar N}^{\left( {Pu} \right)}}(x,t) - \varepsilon _c^{eff\left( 5 \right)}n(x,t){{\bar N}^{\left( 5 \right)}}(x,t).
\label{5.9}
\end{multline}

The same scaling may be applied for the rest 17 equations, eventually yielding:

\begin{eqnarray}
\frac{{\partial \,{N_8}\left( {x,t} \right)}}{{\partial \,t}} &=&  - \,n\left( {x,t} \right)\,\,{N_8}\left( {x,t} \right),\label{5.10}\\
\frac{{\partial \,{N_9}\left( {x,t} \right)}}{{\partial \,t}} &=& n\left( {x,t} \right)\,\left[ {\,{N_8}\left( {x,t} \right) - \varepsilon _c^{\left( 9 \right)}\,{N_9}\left( {x,t} \right)} \right]\, - \varepsilon {N_9}\left( {x,t} \right),\label{5.11}\\
\frac{{\partial \,{N_{Pu}}\left( {x,t} \right)}}{{\partial \,t}}& =& \varepsilon {N_9}\left( {x,t} \right) - \,\left( {\varepsilon _f^{\left( {Pu} \right)} + \varepsilon _c^{\left( {Pu} \right)}} \right)\,\,n\left( {x,t} \right)\,{N_{Pu}}\left( {x,t} \right),\label{5.12}\\
\frac{{\partial \tilde N_i^{\left( {Pu} \right)}(x,t)}}{{\partial {\kern 1pt} t}} &=& \varepsilon _f^{\left( {Pu} \right)}p_i^{\left( {Pu} \right)}n{\kern 1pt} \,\left( {x,t} \right){\kern 1pt} \,{\kern 1pt} {N_{Pu}}\left( {x,t} \right) - \beta _i^{\left( {Pu} \right)}\tilde N_i^{\left( {Pu} \right)}\left( {x,t} \right) - \nonumber\\
&& - \varepsilon _c^{eff\left( {Pu} \right)}n\left( {x,t} \right)\tilde N_i^{\left( {Pu} \right)}\left( {x,t} \right),\quad i = 1,...,6\label{5.13},\\
\frac{{\partial \tilde N_i^{\left( 5 \right)}\left( {x,t} \right)}}{{\partial {\kern 1pt} t}} &=& {\kern 1pt} \varepsilon _f^{\left( 5 \right)}p_i^{\left( 5 \right)}n{\kern 1pt} \,\left( {x,t} \right){\kern 1pt} \,{N_5}\left( {x,t} \right) - \beta _i^{\left( 5 \right)}\tilde N_i^{\left( 5 \right)}(x,t) -\nonumber \\
&& - \varepsilon _c^{eff\left( 5 \right)}n\left( {x,t} \right)\tilde N_i^{\left( 5 \right)}\left( {x,t} \right),\quad i = 1,...,6\label{5.14}
\end{eqnarray}

\begin{eqnarray}
\frac{{\partial {{\bar N}^{\left( {Pu} \right)}}\left( {x,t} \right)}}{{\partial t}} &=& \varepsilon _f^{\left( {Pu} \right)}2\left( {1 - {p^{\left( {Pu} \right)}}} \right)n(x,t){N_{Pu}}(x,t) +\nonumber \\
&& + \sum\limits_{i = 1}^6 {\beta _i^{\left( {Pu} \right)}\tilde N_i^{\left( {Pu} \right)}(x,t) - \varepsilon _c^{eff\left( {Pu} \right)}n\left( {x,t} \right){{\bar N}^{\left( {Pu} \right)}}\left( {x,t} \right)},\label{5.15} \\
\frac{{\partial {{\bar N}^{\left( 5 \right)}}\left( {x,t} \right)}}{{\partial t}} &=& \varepsilon _f^{\left( 5 \right)}2\left( {1 - {p^{\left( 5 \right)}}} \right)n(x,t){N_5}(x,t) + \nonumber\\
&& + \sum\limits_{i = 1}^6 {\beta _i^{\left( 5 \right)}\tilde N_i^{\left( 5 \right)}(x,t) - \varepsilon _c^{eff\left( 5 \right)}n\left( {x,t} \right){{\bar N}^{\left( 5 \right)}}\left( {x,t} \right)}.\label{5.16} 
\end{eqnarray}

In these equations

\begin{equation}
\varepsilon  \equiv \frac{1}{{{V_n}\sigma _c^{\left( 8 \right)}{N_{8,0}}}}\frac{1}{{{\tau _\beta }}}.
\label{5.17}
\end{equation}

Let us set the boundary and initial conditions for these equations. The space
is initially filled with $^{238}U$ and $^{235}U$. So the initial conditions for
them are:

\begin{align}
{\left. {{N_8}\left( {x,t} \right)} \right|_{t = 0}}& = {\eta _8},&{\left. {{N_5}\left( {x,t} \right)} \right|_{t = 0}} &= {\eta _5},
\label{5.18}
\end{align}

\noindent
where $\eta _8$ and $\eta _5$ are constants adjusting the fuel enrichment
($\eta _8+\eta _5=1$).

The remaining elements are initially zero, and their boundary conditions:

\begin{align}
{\left. {{N_9}\left( {x,t} \right)} \right|_{t = 0}} &= 0,&{\left. {{N_{Pu}}\left( {x,t} \right)} \right|_{t = 0}} &= 0,&{\left. {\tilde N_i^{\left( {Pu} \right)}\left( {x,t} \right)} \right|_{t = 0}} &= 0,\\
{\left. {\tilde N_i^{\left( 5 \right)}\left( {x,t} \right)} \right|_{t = 0}} &= 0,&{\left. {{{\bar N}_{Pu}}\left( {x,t} \right)} \right|_{t = 0}} &= 0,&{\left. {{{\bar N}_5}\left( {x,t} \right)} \right|_{t = 0}} &= 0.
\label{5.19}
\end{align}

The initial and boundary conditions for neutrons are

\begin{align}
{\left. {n\left( {x,t} \right)} \right|_{t = 0}} &= 0,&{\left. {n\left( {x,t} \right)} \right|_{x = 0}} &= {n_0}\left( t \right),
\label{5.20}
\end{align}

\noindent
where ${n_0}\left( t \right)$ is some function specifying the number of
external neutrons.

For the numerical simulation we use Wolfram Mathematica. We perform the
calculations for the energy of neutrons $E_n=5.923$~(eV). Unfortunately,
this system currently cannot be solved with real parameters, so let us take
some approximation. We take $\varepsilon = 10^{-3}$. This value is not
realistic ($\varepsilon$ is actually about $10^{-14}$), but it is difficult to
carry out the numerical calculations with more precise values for our model.
The rest of the constants are given below:

\begin{eqnarray}
&&{N_{8,0}} = \frac{{{\rho _8}}}{{{\mu _8}}}{N_A} = \frac{{19}}{{238}}{N_A}\left[ {\frac{1}{{c{m^3}}}} \right];{V_n} = 2.3957 \cdot {10^6}\left[ {\frac{{cm}}{s}} \right];\nonumber\\
&&{\nu _{Pu}} = 2.84;{\nu _5} = 2.38;{\tau _\beta } = 3.3days;\nonumber\\
&&\sigma _f^{\left( {Pu} \right)} = 7.18\left[ b \right];\sigma _c^{\left( {Pu} \right)} = 1.57\left[ b \right];\sigma _c^{\left( 8 \right)} = 6.47\left[ b \right];\sigma _c^{\left( 9 \right)} = 12.95\left[ b \right];\nonumber\\
&&\sigma _f^{\left( 5 \right)} = 7.18\left[ b \right];\sigma _c^{\left( 5 \right)} = 7.18\left[ b \right];{\eta _8} = 0.9;{\eta _5} = 0.1;\nonumber\\
&&\sigma _c^{eff\left( {Pu} \right)} = 011\left[ b \right];\sigma _c^{i\left( {Pu} \right)} = 1.0\left[ b \right],i = 1 \ldots 6;\nonumber\\
&&\sigma _c^{eff\left( 5 \right)} = 0.11\left[ b \right];\sigma _c^{i\left( 5 \right)} = 1.0\left[ b \right],i = 1 \ldots 6;\nonumber\\
&&p_1^{\left( {Pu} \right)} = 0.0072 \cdot {10^{ - 3}};p_2^{\left( {Pu} \right)} = 0.626 \cdot {10^{ - 3}};p_3^{\left( {Pu} \right)} = 0.444 \cdot {10^{ - 3}};\nonumber\\
&&p_4^{\left( {Pu} \right)} = 0.685 \cdot {10^{ - 3}};p_5^{\left( {Pu} \right)} = 0.0072 \cdot {10^{ - 3}};p_6^{\left( {Pu} \right)} = 0.0072 \cdot {10^{ - 3}};\\
&&{p^{\left( {Pu} \right)}} = \sum\limits_{i = 1}^6 {p_i^{\left( {Pu} \right)} = 0.0021} ;T_1^{\left( {Pu} \right)} = 56.28\left[ s \right];T_2^{\left( {Pu} \right)} = 23.04\left[ s \right];\nonumber\\
&&T_3^{\left( {Pu} \right)} = 5.6\left[ s \right];T_4^{\left( {Pu} \right)} = 2.13\left[ s \right];T_5^{\left( {Pu} \right)} = 0.62\left[ s \right];T_6^{\left( {Pu} \right)} = 0.26\left[ s \right];\nonumber\\
&&p_1^{\left( 5 \right)} = 0.21 \cdot {10^{ - 3}};p_2^{\left( 5 \right)} = 1.4 \cdot {10^{ - 3}};p_3^{\left( 5 \right)} = 1.26 \cdot {10^{ - 3}};\nonumber\\
&&p_4^{\left( 5 \right)} = 2.52 \cdot {10^{ - 3}};p_5^{\left( 5 \right)} = 0.74 \cdot {10^{ - 3}};p_6^{\left( 5 \right)} = 0.27 \cdot {10^{ - 3}};\nonumber\\
&&{p^{\left( 5 \right)}} = \sum\limits_{i = 1}^6 {p_i^{\left( 5 \right)} = 0.0064} ;T_1^{\left( 5 \right)} = 55.72\left[ s \right];T_2^{\left( 5 \right)} = 22.72\left[ s \right];\nonumber\\
&&T_3^{\left( 5 \right)} = 6.22\left[ s \right];T_4^{\left( 5 \right)} = 2.3\left[ s \right];T_5^{\left( 5 \right)} = 0.61\left[ s \right];T_6^{\left( 5 \right)} = 0.23\left[ s \right];\nonumber
\label{5.21}
\end{eqnarray}

A numerical simulation of the wave kinetics was performed for the case of
persistent source of external neutrons, as well as for the case of switching
the source off after the wave burning is established. Switching off the
external source of neutrons allows to test the autowave mode.

When simulating the case of neutron source switching off, the function of
external neutrons was given as:

\begin{equation}
n_0(t)=3000 \cdot t \cdot \exp \left(-0.5 \cdot t \right).
\label{5.22}
\end{equation}

\begin{figure}
\begin{center}
\includegraphics[width=0.9\linewidth]{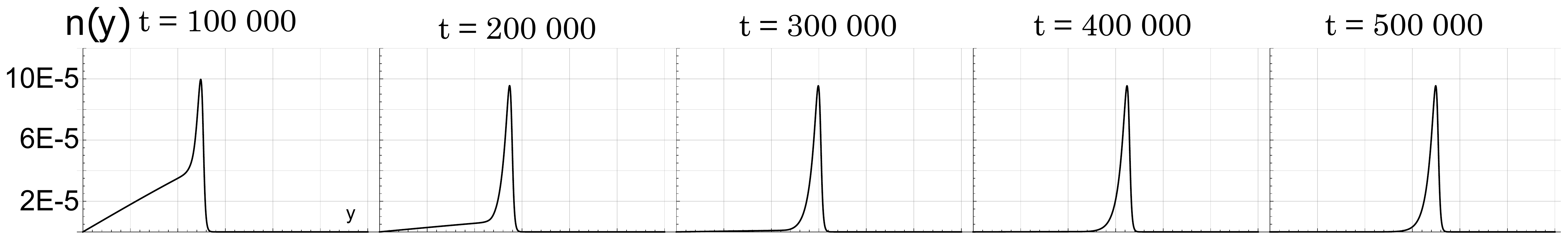}
\includegraphics[width=0.9\linewidth]{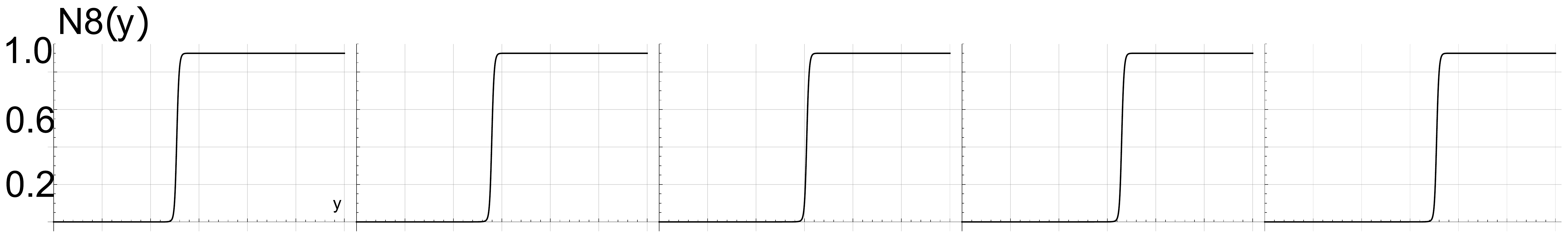}
\includegraphics[width=0.9\linewidth]{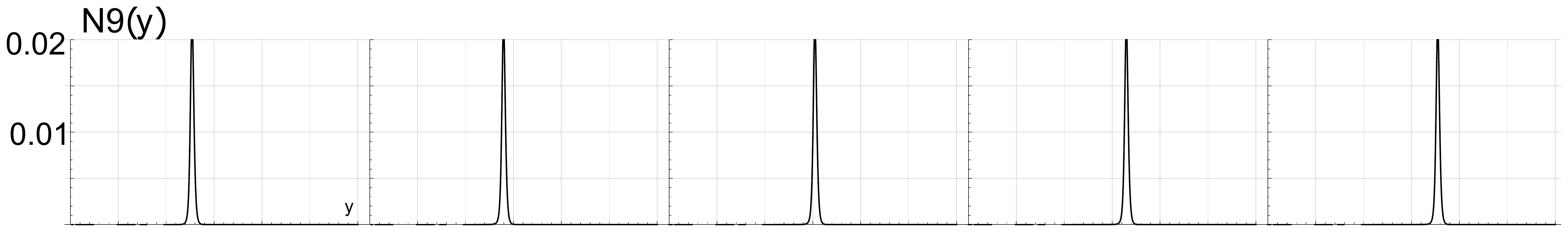}
\includegraphics[width=0.9\linewidth]{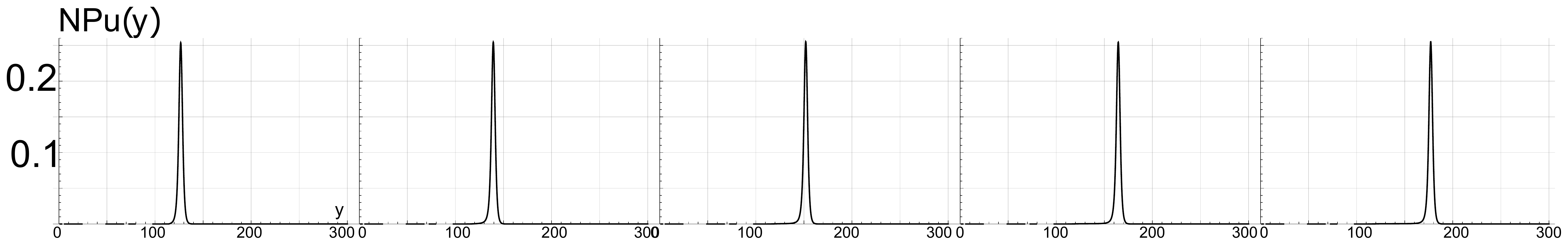}
\end{center}
\caption{The concentrations of neutrons and nuclides in time from $t=100~000$
to $t=500~000$.}
\label{fig5.1}
\end{figure}

Fig.~\ref{fig5.1} presents the results of numerical simulation (the neutron
energy was $E_n=5.923$~(eV)) with the external source of neutrons being
switched off at some point. The setting of the autowave mode of nuclear burning
is confirmed by the fact that starting from a certain moment in time
(when the external source of neutrons is already switched off), each subsequent graph
differs from the previous only by a shift along the $y$-axis.

We also studied the regions of neutron energies for which there was no
minimum in potential energy. In such cases the autowave mode did not establish,
and the wave faded away after the external source of neutrons was switched off.

\section{Conclusions}

A kinetic system of equations, describing the wave mode of neutron-nuclear
burning in uranium-plutonium medium, is formulated. Its autowave form is also
obtained. In contrast to many other papers like~\citep{9,15,38,39,40,41}, we do
not neglect the derivative of neutron concentration with respect to time in the
neutron diffusion equation. This way we study the non-stationary burning mode.

For the first time a kinetic equation for neutrons is obtained in the form of
the energy conservation law for a mechanical system with dissipation, and the 
mechanical analogy for the fission wave is developed. This allowed to formulate
the conditions for the existence of the autowave burning mode, and determine
the possible values of the wave speed. We also succeeded in determining the
regions of neutron energy, for which the autowave burning is possible. We
conclude that in other energy ranges, in which the autowave mode is impossible,
the wave of nuclear burning may still be established using the support of an
external neutron source.

To confirm the theoretical conclusions, we performed a numerical 1D simulation
of the neutron-nuclear burning in uranium-plutonium medium in a single-group
diffusion approximation ($E_n = 5.923$~(eV)). The results of numerical
modelling confirm the obtained theoretical conclusions. According to these
results, one of the possible areas of autowave burning in the uranium-plutonium
medium is the epithermal region of neutron energies (as in~\citep{32,42}).
Meanwhile in the region of fast neutrons, the wave burning of the
uranium-plutonium medium requires the constant supply of neutrons from an
external source, and the wave fades out when the source is switched off.

The discovered burning mode with external support can obviously be used for
implementing a traveling-wave reactor, e.g. to stop the burning at any time
by switching off the external source of neutrons.



\end{document}